\documentclass[lettersize,journal]{IEEEtran}
\usepackage{amsmath,amsfonts}
\usepackage{algorithmic}
\usepackage{algorithm}
\usepackage{array}
\usepackage[caption=false,font=normalsize,labelfont=sf,textfont=sf]{subfig}
\usepackage{textcomp}
\usepackage{stfloats}
\usepackage{url}
\usepackage{verbatim}
\usepackage{graphicx}
\usepackage{cite}
\begin{document}

\title{Spatiotemporal computing for temporal nonlinear tasks using few-mode fiber and SOA nonlinearity}

\author{Adhila Nazarudeen, Silvia Ortín, Apostolos Argyris
        % <-this % stops a space
\thanks{Adhila Nazarudeen and Apostolos Argyris are with the Institute for Cross-Disciplinary Physics and Complex Systems - IFISC (UIB-CSIC), 07122 Palma de Mallorca, Spain (e-mail: adhila@ifisc.uib-csic.es; apostolos@ifisc.uib-csic.es).}
\thanks{Silvia Ortín was with the Institute for Cross-Disciplinary Physics and Complex Systems - IFISC (UIB-CSIC), 07122 Palma de Mallorca, Spain, when this work was performed. She is now with the Institute of Physics of Cantabria (IFCA), CSIC-University of Cantabria, 39005 Santander, Spain (e-mail: ortin@ifca.es).}% <-this % stops a space
\thanks{This work was funded by the European Commission under the Horizon Europe grant 101169118 (POST-DIGITAL+). It was also partially funded by the Program for Centres and Units of Excellence in R\&D, María de Maeztu under Project CEX2021-001164-M/10.13039/501100011033.}}

\markboth{This work has been submitted to the IEEE for possible publication. Copyright may be transferred without notice.}
{Shell \MakeLowercase{\textit{et al.}}: A Sample Article Using IEEEtran.cls for IEEE Journals}

\IEEEpubid{}

\maketitle

\begin{abstract}
Step-index few-mode fibers (FMFs) provide a compact passive platform for high-speed photonic information processing by exploiting modal dispersion to map temporal inputs into spatiotemporal representations with short-term memory. 
In previous studies, we have demonstrated linear classification tasks with such FMFs, such as ultrafast multibit header recognition. 
Here, we extend the computational capability of such architectures towards solving nonlinear tasks by introducing an optical nonlinearity at the output of the dispersive medium. 
The FMF output is processed through parallel direct and nonlinear branches, with the latter implemented using a nonlinear semiconductor optical amplifier (SOA). 
We experimentally evaluate the architecture at 28.5 Gb/s on two delayed XOR and higher-order parity tasks. 
We show that combining the direct and SOA-transformed representations reduces classification errors and broadens the range of operating conditions supporting low-error classification.
In addition, by exploiting multiple temporal samples within each bit period as independent classifier features, we increase the dimensionality of the photonic representation and improve the computational performance.

\end{abstract}


\section{Introduction}

\IEEEPARstart{P}{hotonic} computing has emerged as a promising approach for performing information processing in the optical domain, leveraging the intrinsic parallelism, the high bandwidth, and light-speed propagation. 
Rather than reproducing the operation of conventional digital processors, photonic systems can exploit propagation, interference, and light--matter interactions to implement physical transformations of information with potentially high throughput and low latency \cite{McMahon2023,Wetzstein2020,Shastri2021,Feldmann2021}. 
Among the different approaches to photonic computing, feed-forward architectures process information through a sequence of physical transformations as light propagates through optical elements or complex optical media \cite{Shen2017,Lin2018,Zhou2021,Xu2021}.
Such architectures are well suited to transformations in which each input can be processed independently.
Many computational problems, however, involve temporally structured information, for which the desired output depends not only on the current input but also on inputs presented at earlier times.
Extending feed-forward photonic computing to this regime therefore requires the optical transformation itself to retain and process information across time.
Several physical mechanisms have been exploited to introduce such temporal memory into photonic computing systems.
In recurrent and reservoir-computing architectures, memory typically emerges from feedback, delayed interactions, or the intrinsic dynamics of nonlinear optical systems, providing the capability to process temporally correlated signals \cite{Appeltant2011, Paquot2012, Brunner2013, Vandoorne2014, VanderSande2017, Harkhoe2019, Brunner2019, Yan2024, Chen2026, Abdalla2026}. 
Temporal memory, however, does not need to rely on recurrence or dynamical nonlinearities; it can also emerge directly during feed-forward propagation when different components of an optical signal accumulate different propagation delays, causing information injected at different times to overlap and interact at the output.

Multimode propagation provides a natural mechanism for realizing this type of processing. 
When multiple spatial modes propagate simultaneously, mode mixing can generate complex spatial transformations, while differences in their propagation characteristics can additionally couple the spatial and temporal degrees of freedom of the optical field \cite{Richardson2013, Wright2015, Xiong2016, Porte2021, Cao2023, Skalli2025}.
As a result, a temporally encoded input can undergo a distributed spatiotemporal transformation during propagation, with the differential modal delays providing access to a finite history of the input signal. 
In this way, the temporal memory is not introduced through an explicit feedback loop, but is embedded in the propagation itself. 
Researchers have exploited multimode mixing and dispersion for information processing across different platforms, including multimode waveguides \cite{Katumba2018, Wu2021, Heinsalu2024}, complex optical media \cite{Matthes2019}, and multimode fibers (MMFs) \cite{Carpenter2014, Rahmani2018, Fan2021, Cao2023}.
The latter are particularly attractive for realizing such propagation-based spatiotemporal transformations.
Their complex modal transformations have been widely exploited for spatial information processing, mapping input optical fields onto high-dimensional output responses \cite{DiLeonardo2011, Carpenter2014, Ploschner2015, Rahmani2018, Borhani2018, Paudel2020, Fan2021, Cao2023}.
FMFs support only a limited number of spatial modes, reducing the complexity of the modal propagation while retaining the multimode linear mixing described above. 
This reduced complexity can facilitate a more controlled exploitation of the propagation characteristics for information processing applications. 
In particular, controlled differential delays among spatial modes have been used to implement true-time-delay lines and microwave-photonic finite-impulse-response filters \cite{Nickel2017, Nazemosadat2020, Nazemosadat2022}. 
More recently, we exploited these propagation characteristics for temporal computing using a step-index FMF operated as a linear dispersive  element \cite{Ortin2022, Pfluger2024}. 
The passive spatiotemporal transformation produced by the FMF provided sufficient temporal memory to perform ultrafast multi-bit temporal classification using a short temporal readout window, establishing the FMF as a compact platform for high-speed spatiotemporal processing.

Building on this platform, the present work extends the temporal processing toward nonlinear computational tasks. 
Nonlinear transformations in MMF computing can be achieved through nonlinear propagation and modal interactions within the fiber itself \cite{Krupa2017, Tegin2021, Wright2022, Yildirim2023, Oguz2024, Kesgin2025, Marcucci2025}. 
An alternative approach is to retain the fiber as a linear dispersive medium and introduce the nonlinear transformation separately. 
Here, we follow this approach by adding a nonlinear semiconductor optical amplifier (SOA) after the FMF, whose gain-saturation response provides an additional nonlinear transformation of the dispersed optical signal. 
SOA gain saturation has previously been exploited in photonic information processing and computing \cite{Duport2012, Manuylovich2024}.
Importantly, this configuration preserves access to the directly detected FMF output while simultaneously generating an additional SOA-transformed representation, allowing both representations of the same temporally dispersed input to be exploited for processing.
We experimentally investigate this dual-representation approach for nonlinear temporal computing, evaluating delayed two-bit XOR and multi-bit parity tasks. 
We assess the contribution of the SOA nonlinear transformation and determine whether combining the two representations improves the processing of information distributed across multiple input bits. 
Finally, we examine how temporal sampling and readout dimensionality affect classification performance.

\begin{figure*}[!t]
\centering
\includegraphics[width=1\textwidth]{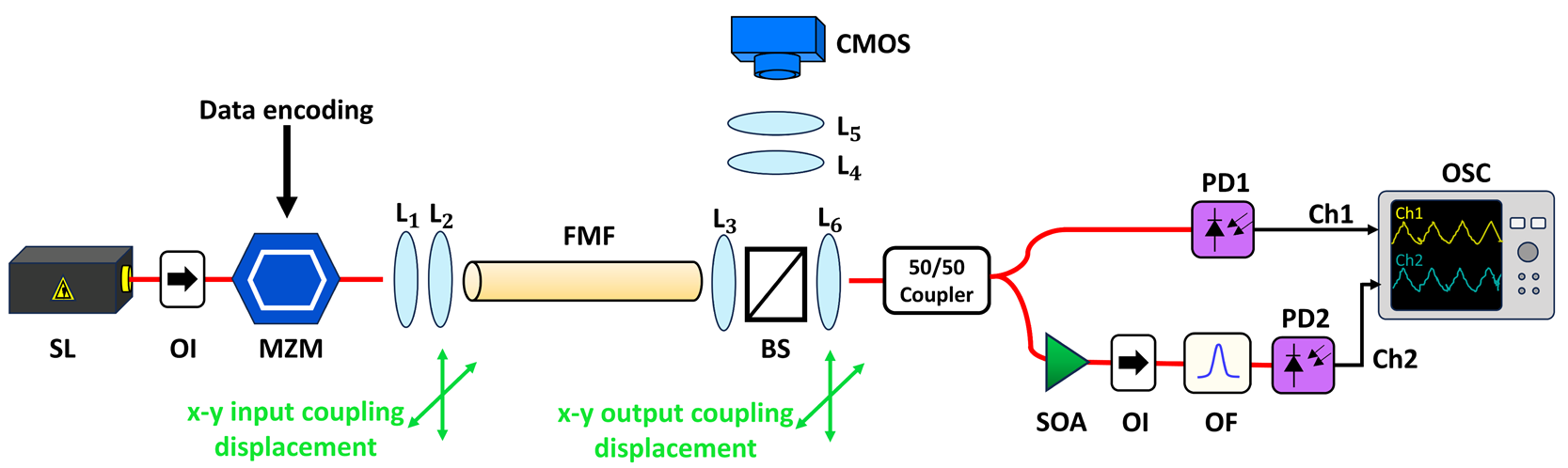}
\caption{Experimental setup. Red lines indicate light propagation in polarization-maintaining single-mode fibers (PM-SMFs). $L_{1}$--$L_{6}$: Focusing and collimation lenses. $L_{2}$ and $L_{6}$ can be moved via piezoelectric translational stages. SL: Semiconductor laser, OI: Optical isolator, MZM: 40 GHz Mach-Zehnder modulator, BS: Beam splitter, CMOS: CMOS camera, SOA: Nonlinear semiconductor optical amplifier, OF: Optical filter, PD: Photoreceiver, OSC: 40 GHz Real-time oscilloscope.}
\label{setup}
\end{figure*}

\section{Experimental Setup and Data Acquisition}
\label{sec:experiment}

\subsection{Experimental setup}

Fig.~\ref{setup} shows the experimental setup, which builds on the FMF-based architecture previously reported in \cite{Pfluger2024}, where a detailed description of the system can be found. The main modification introduced in the present work is the addition of a second, SOA-based output branch in parallel with the existing direct detection path. The resulting dual-branch configuration enables simultaneous acquisition of the FMF response through the original direct photodetection path and through a second path that introduces an additional nonlinear optical transformation before photodetection.

\begin{figure}[!t]
\centering
\includegraphics[width=\linewidth]{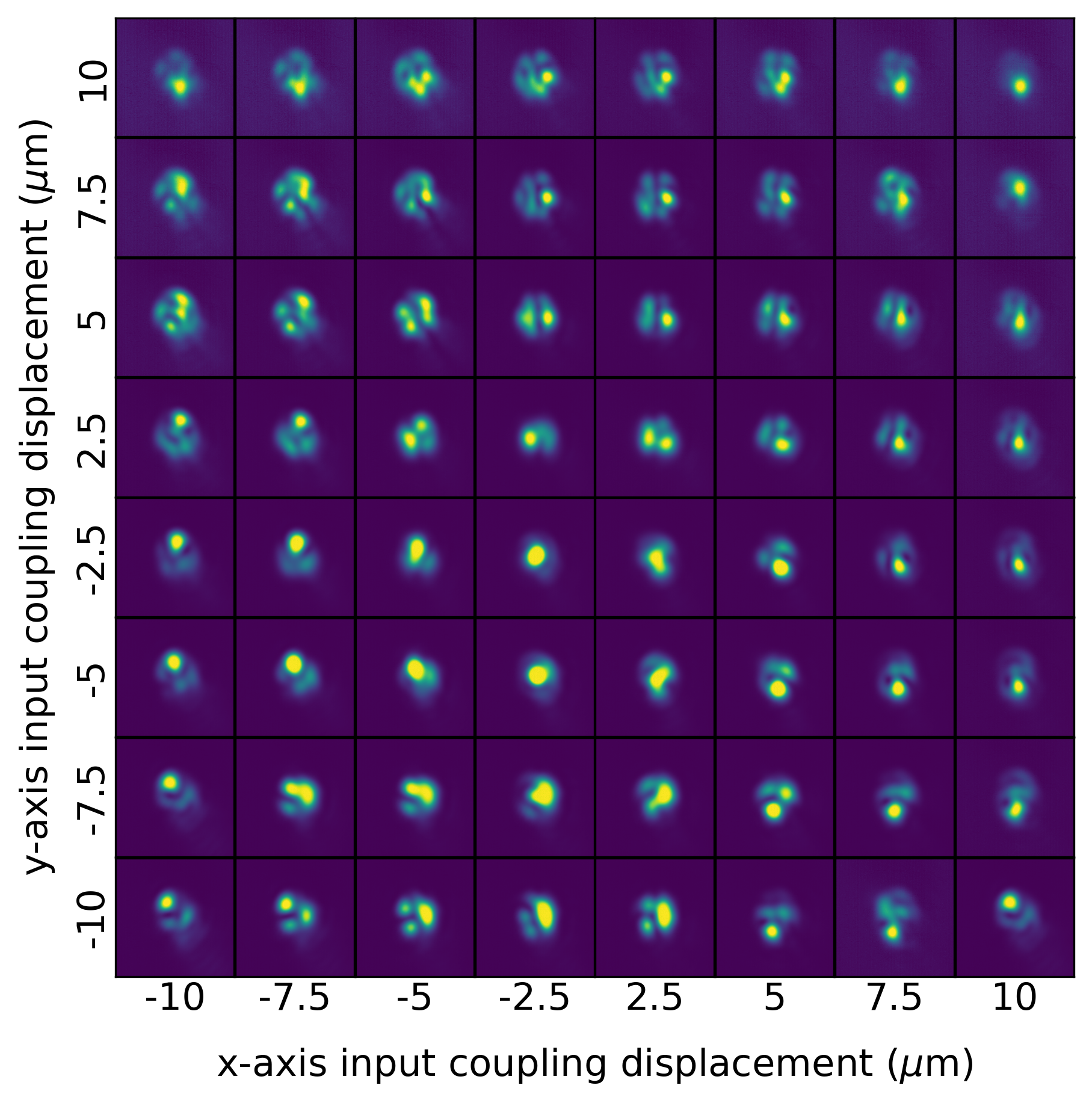}
\caption{Interference patterns captured by the CMOS camera at the output of the FMF for each one of the 64 SMF-FMF coupling conditions defined by the input coupling displacement along the $X$ and $Y$ axes.}
\label{grid}
\end{figure}

Briefly, a continuous-wave optical carrier at 1550 nm is intensity-modulated at 28.5 Gb/s and coupled into a 13 m step-index FMF (22 µm core diameter, NA = 0.15) that supports 12 linearly polarized (LP) modes. 
The transverse SMF--FMF coupling position is controlled with a piezoelectric translation stage. 
By tuning the input coupling displacement in two axes, we can excite different combinations of spatial modes and thus different patterns of temporal linear mixing. 
In this work, we evaluate 64 SMF-FMF input coupling conditions distributed over a two-dimensional $8\times8$ grid, with a displacement range of [-10 $\mu$m, 10 $\mu$m] from the central alignment.
To monitor the interference patterns at the output of the FMF, a CMOS camera is employed. 
The output intensity distributions obtained for the 64 input coupling conditions are shown in Fig.~\ref{grid}, illustrating the different spatial interference patterns generated as the SMF--FMF coupling position is varied.

\begin{figure}[!h]
\centering
\includegraphics[width=0.7\linewidth]{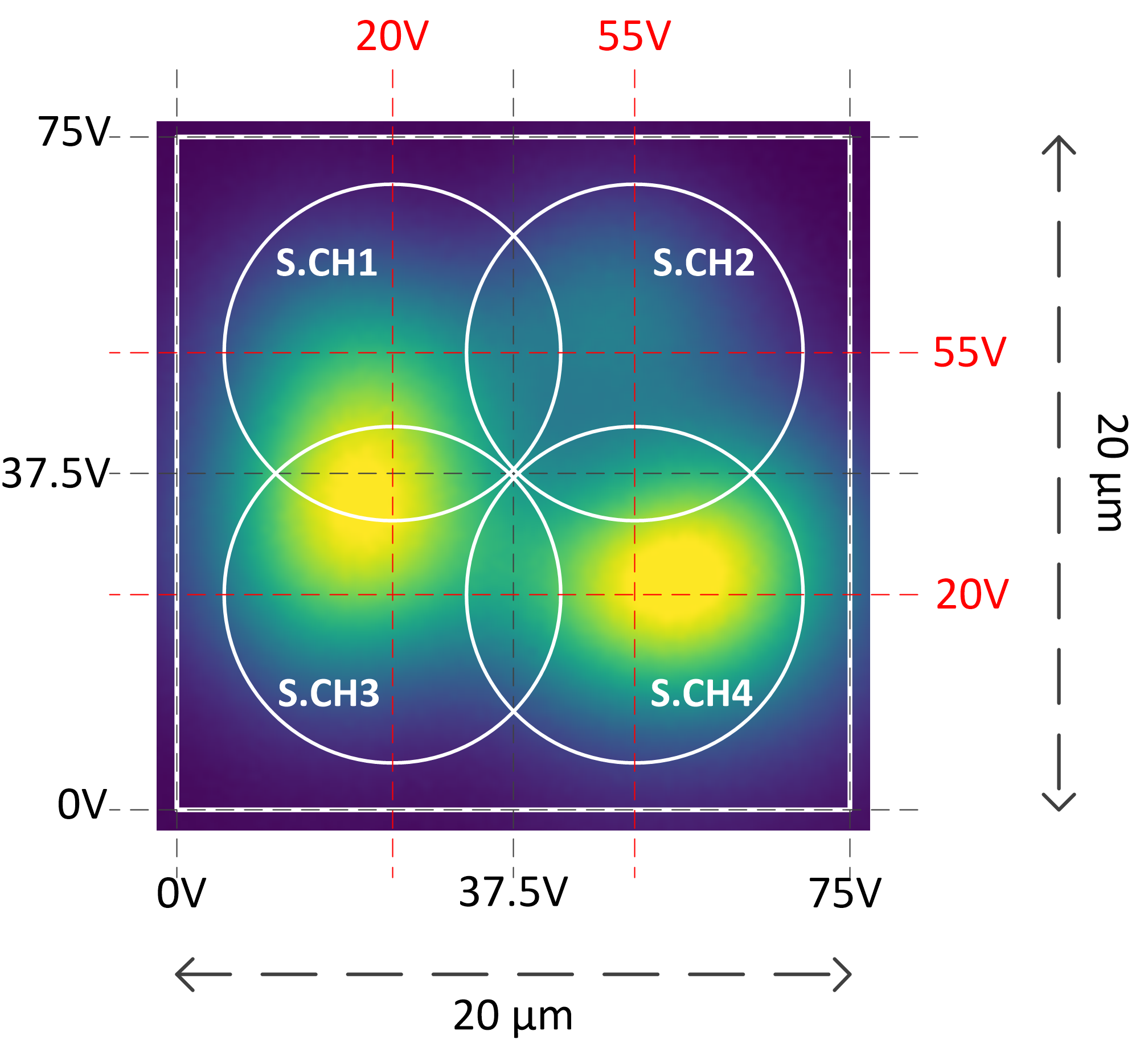}
\caption{Allocation of the four quadrant positions of the FMF--SMF coupling over the spatial mode distribution pattern. The four white circles (S.CH1--4) indicate the approximate regions that collimate the optical power into the output SMF, assuming a field mode diameter of 10~$\mu$m. The 0–75 V range applied to each axis of the piezoelectric stage corresponds to a total displacement of approximately 20~$\mu$m, according to the device specifications. The example corresponds to the case of (2.5 $\mu$m, 2.5 $\mu$m) input coupling displacement.}
\label{speckle}
\end{figure}

To acquire both spatial and temporal information from the FMF output without spatially integrating the entire output field into a single detected signal, the output coupling SMF is sequentially displaced to four different positions in the image plane (see Fig.~\ref{speckle}). 
At each position, a temporal waveform is recorded, providing a different spatial projection of the modal interference pattern.
The four output spatial positions are selected using a second piezoelectric translation stage that controls the FMF--SMF output coupling.
To initialize the alignment, we first set a zero input-displacement condition at lens $L_{2}$, corresponding to the position that maximizes the optical power coupled into the FMF. 
We then adjust the output displacement at $L_{6}$ to the position that maximizes coupling into the output SMF.
The coupled optical power is monitored using a power meter. 
This optimized central condition is defined at 37.5 V in each axis of the piezoelectric mount and is the center of our output grid (Fig.~\ref{speckle}).
By estimating a field mode diameter of 10 $\mu$m of the output SMF, we set the voltages of the x and y axes to 20 V or 55 V in order to define the four spatial channels. 
These voltages define four output spatial positions that collectively cover the four quadrants of the optical power distribution at the FMF output, as shown with the white circles in Fig.~\ref{speckle}.
Since these four spatial channels cannot be acquired simultaneously with the present setup, the same input sequence is repeated for each output spatial position.

\begin{figure}[!t]
\centering
\includegraphics[width=\linewidth]{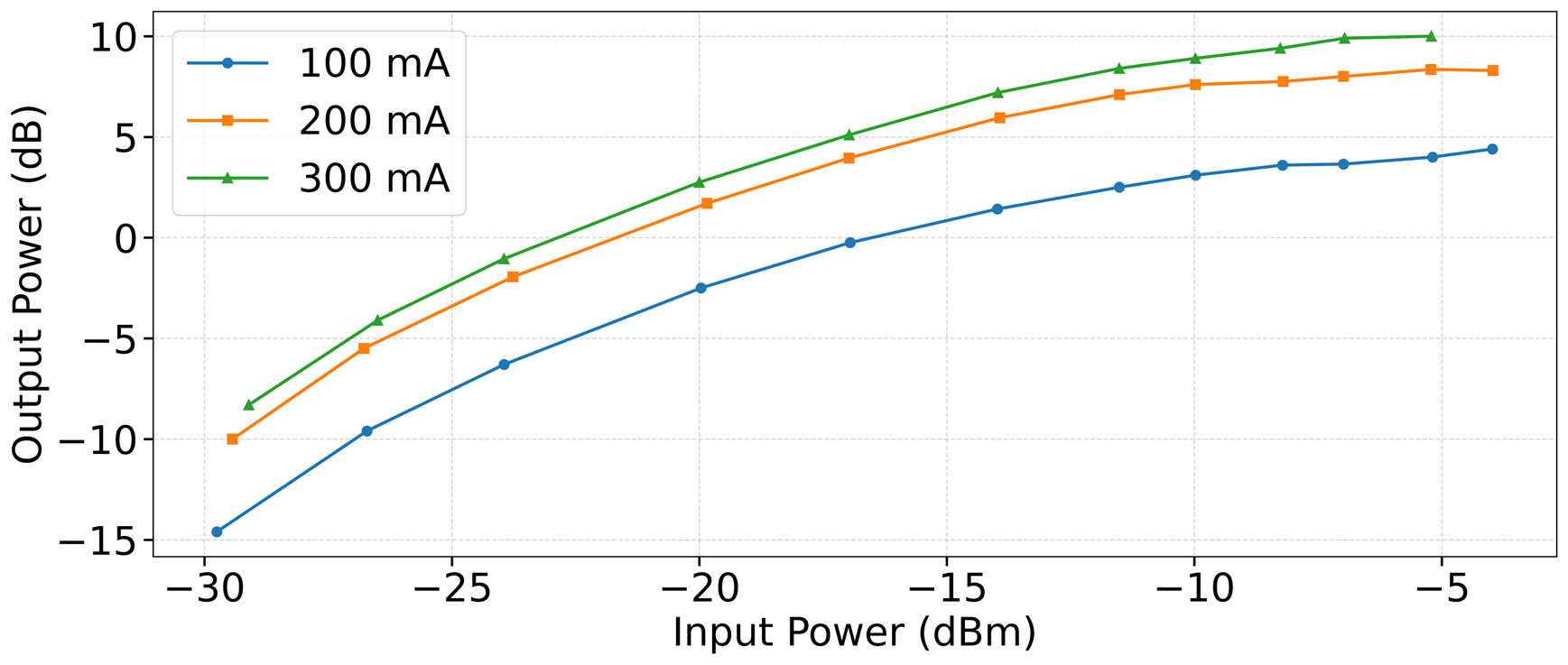}
\caption{Nonlinear SOA transfer function characterization for different bias currents.}
\label{transfer}
\end{figure}

In this study, for each output spatial position, the collected temporal signal is split into two paths using a 50/50 coupler. 
One path is directly photodetected, whereas in the second path the optical signal passes through a nonlinear SOA (model CIP NL-OEC-1550).
The nonlinear response of the SOA was characterized by measuring its input-output optical power transfer function for different bias currents, as shown in Fig.~\ref{transfer}.
It exhibits gain saturation even at input optical powers as low as -25 dBm, becoming progressively more pronounced as the input power increases.
For the experiments reported in this study, the bias current of the SOA was 200 mA. 
Due to the different modal excitation conditions and spatial sampling positions, the optical power entering the SOA varies across the acquired spatiotemporal channels, resulting in different operating points along its nonlinear transfer function.
The low gain saturation level guarantees that even the low-power signals undergo a nonlinear transformation in this branch.
An optical isolator minimizes the reflections that may back-propagate through the SOA and affect its gain temporal response. 
A Gaussian optical filter with a 1 nm 3-dB bandwidth is placed after the SOA to suppress the amplified spontaneous emission noise before the signal is detected by the second 40 GHz photoreceiver (Fig.~\ref{setup}).
Finally, the direct and SOA-transformed signals are recorded simultaneously with a real-time oscilloscope (Keysight UXR0404A, 256 GSa/s). 
Repeating the data acquisition sequentially for the four output spatial conditions, we obtain four spatially resolved temporal traces per branch, and in total eight temporal traces for a given input coupling condition.

\subsection{Temporal encoding and data acquisition}

\begin{figure}[!t]
\centering
\includegraphics[width=\linewidth]{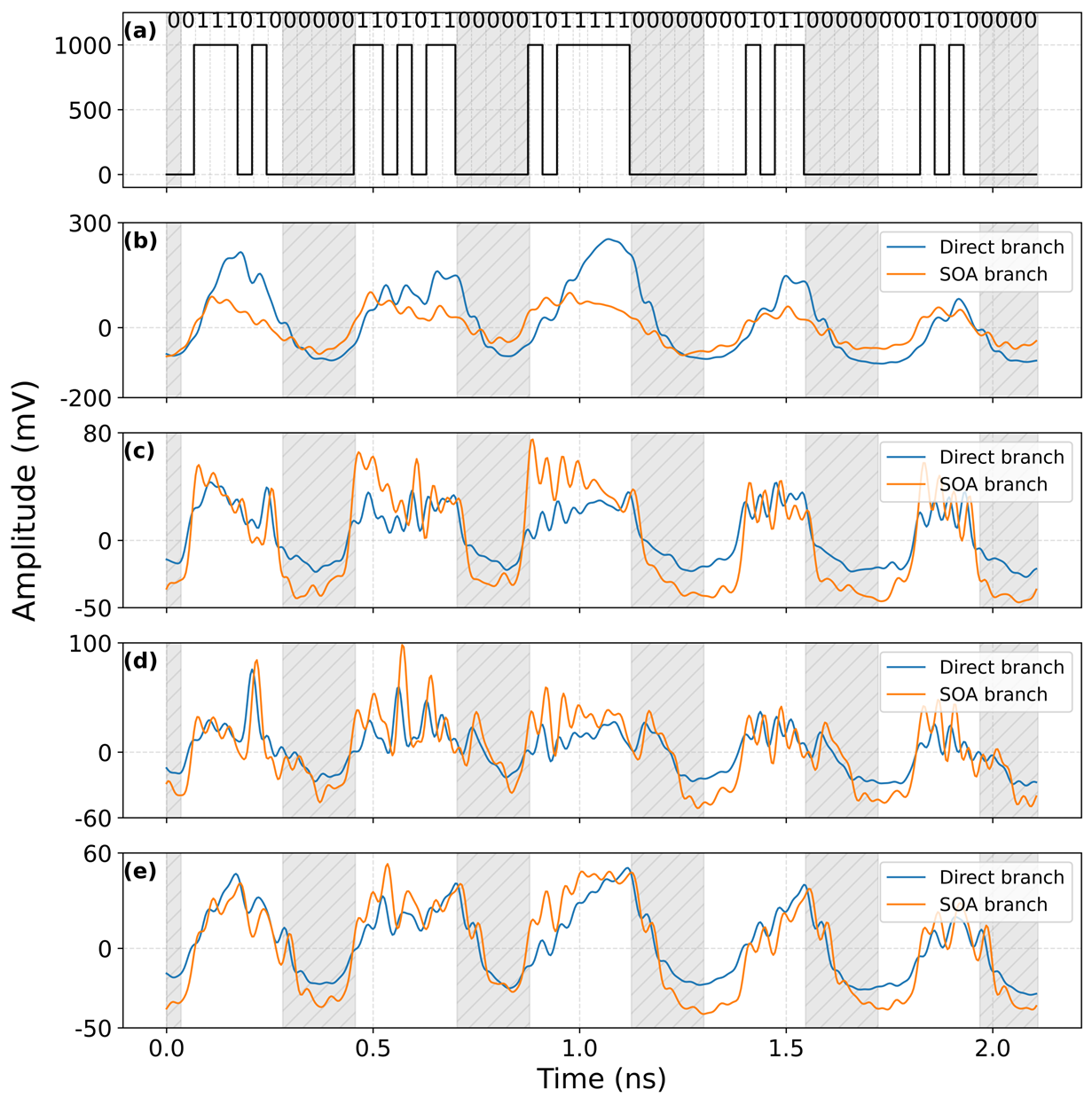}
\caption{
Example of the encoded temporal input and measured responses of the dual-branch system.
(a) OOK-encoded input, consisting of consecutive 7-bit sequences separated by five-zero-bit buffers (shaded regions).
(b)--(e) Measured temporal responses at the four output spatial positions (S.CH1--4).
In each panel, the direct and nonlinear SOA responses correspond to the same output spatial position and input sequence.
}
\label{fig_2}
\end{figure}

The temporal input consists of a series of 7-bit binary sequences encoded using on--off keying (OOK), where the presence or absence of an optical pulse represents a bit value of 1 or 0, respectively. 
Each bit is encoded using three temporal samples generated by an arbitrary waveform generator operating at $85.3\overline{3}$ GSa/s.
Each 7-bit sequence is followed by five consecutive zero bits, which provide temporal separation between successive sequences and suppress inter-sequence crosstalk. 
The system response is acquired with a real-time oscilloscope at a sampling rate of 256~GSa/s, resulting in 9 temporal samples per bit.
For each input coupling condition, the same set of 7-bit sequences is acquired at the four output spatial positions described in the previous subsection, resulting in four temporal traces from each branch. 
The four direct and four SOA traces constitute the eight spatiotemporal channels used for the combined Direct+SOA readout. 
An example of temporal responses from the two branches at the four output spatial positions is shown in Fig.~\ref{fig_2}.

\section{Nonlinear Benchmark Tasks and Feature Extraction}
\label{sec:task}

The computational capability of this system is evaluated using parity-type temporal tasks, which require nonlinear combinations of selected bits within each 7-bit input sequence. 
Let \(x_k\in\{0,1\}\) denote the value of the \(k\)-th bit in the sequence. 
We consider two delayed two-bit XOR tasks:
\[
y_{7,4}=x_7\oplus x_4, \qquad
y_{7,3}=x_7\oplus x_3,
\]
where \(\oplus\) denotes the XOR operation, and one three-bit parity task:
\[
y_{7,6,5}=x_7\oplus x_6\oplus x_5.
\]

The two-bit XOR tasks involve the same nonlinear operation but differ in the temporal separation between the relevant bits, allowing the effect of temporal memory to be probed under comparable nonlinear requirements. 
In contrast, the parity(7–6–5) task requires the nonlinear combination of three temporally adjacent bits. 
All these tasks provide benchmarks with different demands on temporal memory and nonlinearity.

For each benchmark task, the dataset contains a total of 7000 7-bit input sequences, divided into 1500 training sequences, 500 validation sequences, and 5000 test sequences.
The dataset is constructed so that all \(2^7=128\) possible 7-bit input sequences are approximately equally represented.
The performance is quantified by the classification error rate. 
Given the test set size, the minimum non-zero measurable classification error is \(1/5000=2\times10^{-4}\). Cases with no misclassified test sequences are reported as error-free over the test set.
A logistic regression classifier without regularization is used to solve the benchmark tasks. 
Its input features are obtained from a temporal readout window spanning one bit duration, corresponding to \(M\) = 9 consecutive temporal samples in the acquired waveform. 
Each retained sample contributes one feature per spatial channel and branch, so that the total number of features, $N_f$, is:

\[
N_f=M\times N_{\mathrm{branches}}\times N_{\mathrm{spatial}}.
\]

Using the full temporal readout resolution (\(M=9\)), it results in \(4\times9=36\) features for a single branch and \(2\times4\times9=72\) features for the combined Direct+SOA readout.

The same temporal readout window is used for all spatial channels and for the Direct, SOA, and combined Direct+SOA readouts. 
For each input coupling condition and benchmark task, the window position is selected by minimizing the validation error of the combined Direct+SOA readout. 
Once selected, the window position is kept fixed, and the classifiers are retrained using the combined training and validation sets, comprising 2000 sequences, before evaluation on the independent 5000-sequence test set.
All features are standardized to zero mean and unit variance using statistics computed from the training set. 
Although the target tasks depend on two or three bits of the input sequence, the classifier only accesses the information contained within a one-bit temporal readout window of the measured response.

\section{Results and Discussion}
\label{sec:results}

\subsection{Performance on nonlinear temporal tasks}
\label{sec:performance}

\begin{figure}[!t]
\centering
\includegraphics[width=\linewidth]{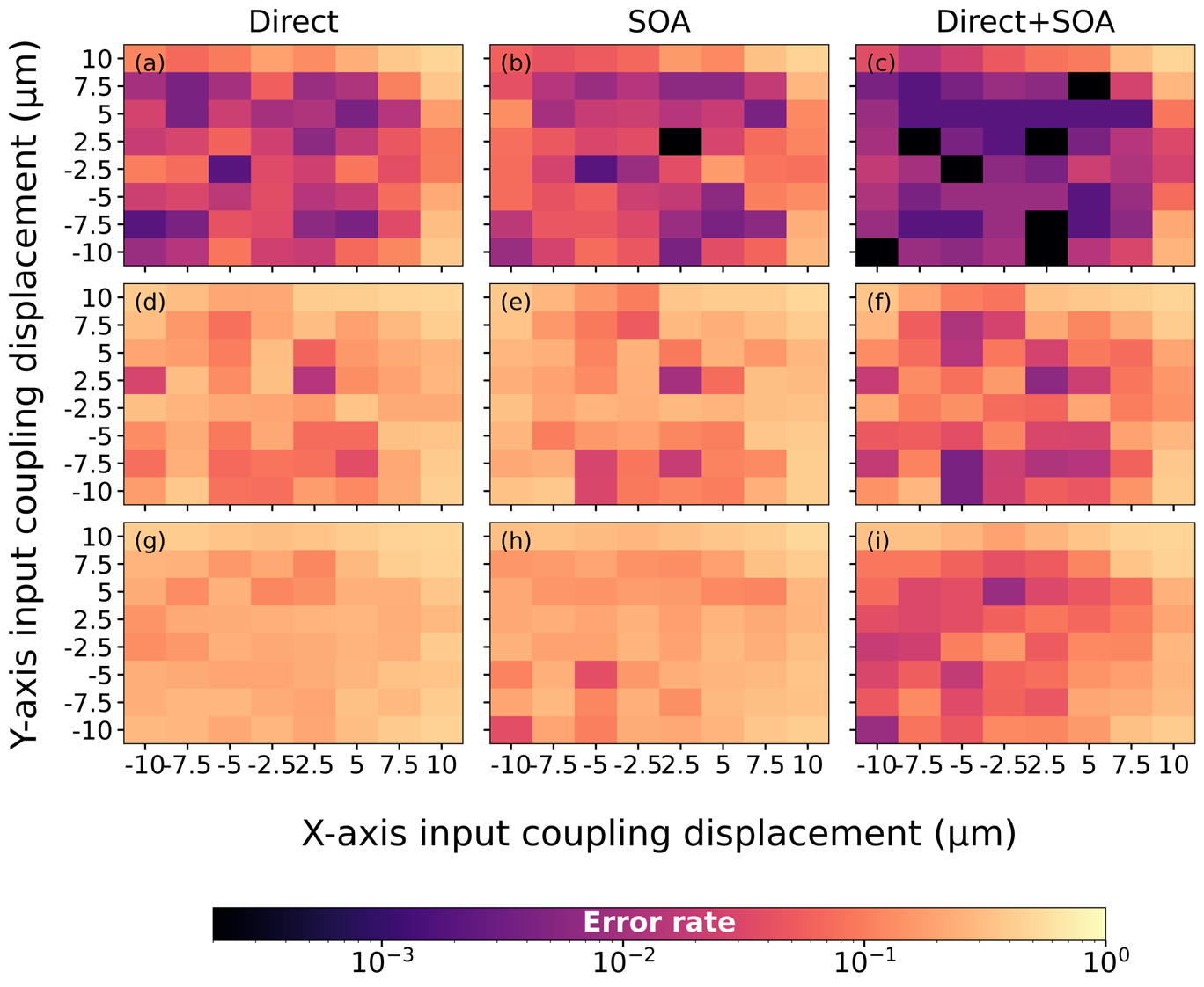}
    \caption{Classification error rate across the 64 input coupling conditions distributed on the \(8\times8\) grid for the three benchmark tasks. 
    Panels (a)--(c) correspond to XOR(7--4), panels (d)--(f) to XOR(7--3), and panels (g)--(i) to the parity(7--6--5) task.
    Each column corresponds to a different readout configuration: Direct, SOA, or combined Direct+SOA. 
    The full temporal readout resolution (\(M=9\)) is used.
}
    \label{fig:branch_comparison}
\end{figure}

We evaluate the computational performance of the ``Direct'', ``SOA'', and combined ``Direct+SOA'' readouts on the three nonlinear temporal tasks using the full temporal readout resolution (\(M=9\)).
Fig.~\ref{fig:branch_comparison} shows the classification error across the 64 input coupling conditions.
Table~\ref{tab:branch_comparison} summarizes the minimum error and the number of conditions achieving an error rate of \(10^{-2}\) or lower.

We first consider only the direct branch, which corresponds to the FMF-based architecture previously reported in \cite{Pfluger2024} and used for temporal linear classification tasks. 
For the XOR(7--4) task, 21.9\% of the input coupling conditions achieve an error rate of \(10^{-2}\) or lower. 
The performance deteriorates for the XOR(7--3) task, for which only 1.6\% of the conditions reach this error level, while none of the conditions do so for the parity(7--6--5) task. 
These results indicate that the temporal mixing provided by the FMF, together with the intrinsic square-law nonlinearity of photodetection, does not consistently solve the two-bit XOR tasks.
It is also not sufficient to solve the three-bit parity task.

If we consider only the nonlinear SOA branch, the number of input coupling conditions achieving low-error operation remains largely unchanged. 
However, this additional nonlinearity improves the minimum classification error across the three tasks. 
The largest improvement is observed for the XOR(7--4) task, with a fivefold reduction in minimum error, while for the XOR(7--3) and the parity(7--6--5) tasks the improvement is by a factor of approximately 1.7 and 2.4, respectively. 
Despite this improvement for the parity task, we cannot identify cases with error rates of \(10^{-2}\).

\begin{table}[!b]
\centering
\caption{Summary of the classification results shown in
Fig.~\ref{fig:branch_comparison}, reporting the minimum error and the number (percentage) of input coupling conditions achieving an error rate of \(10^{-2}\) or less.}
\label{tab:branch_comparison}
\begin{tabular}{llcc}
\hline
Readout & Task & Min. error & \# Conditions $\leq10^{-2}$ \\
\hline
Direct
 & XOR(7--4)       & $2.0\times10^{-3}$ & 14 (21.9\%) \\
 & XOR(7--3)       & $1.0\times10^{-2}$ & 1 (1.6\%) \\
 & Parity(7--6--5) & $9.0\times10^{-2}$ & 0 (0\%) \\
\hline
SOA
 & XOR(7--4)       & $4.0\times10^{-4}$ & 17 (26.6\%) \\
 & XOR(7--3)       & $6.0\times10^{-3}$ & 1 (1.6\%) \\
 & Parity(7--6--5) & $3.8\times10^{-2}$ & 0 (0\%) \\
\hline
Direct+SOA
 & XOR(7--4)       & $\mathbf{2\times10^{-4}}$ & \textbf{40 (62.5\%)} \\
 & XOR(7--3)       & $\mathbf{4.0\times10^{-3}}$ & \textbf{7 (10.9\%)} \\
 & Parity(7--6--5) & $\mathbf{8.0\times10^{-3}}$ & \textbf{3 (4.7\%)} \\
\hline
\end{tabular}
\end{table}

However, when considering the combined Direct+SOA readout, in which features from both branches are jointly processed, the most substantial improvement is seen in the number of input coupling conditions that achieve low-error operation. 
For the XOR(7--4) task, 62.5\% of the conditions reach an error rate of \(10^{-2}\) or lower, compared with 21.9\% and 26.6\% for the direct and SOA branches, respectively. 
For the XOR(7--3) task, this fraction increases to 10.9\%, compared with only 1.6\% for either individual branch. 
For the parity(7--6--5) task, neither of the individual branches reaches the \(10^{-2}\) error level, whereas 4.7\% of the conditions do so with the combined readout. 
In terms of minimum error, a further reduction in classification error is observed across all tasks.

Overall, the results highlight distinct effects of the SOA nonlinear transformation and the combination of the two readout branches. 
The SOA branch consistently improves the best achievable performance relative to the direct readout, while having only a limited effect on the fraction of input coupling conditions that yield low error. 
In contrast, the combined Direct+SOA readout substantially increases this fraction across all three tasks while also achieving the lowest minimum errors. 
However, this improvement is accompanied by a doubling of the number of features, from 36 to 72. 
It therefore remains unclear whether the improved performance results from jointly exploiting the two physical representations or simply from the increased feature dimensionality. 
This question is examined in the following subsection.

\subsection{Effect of feature dimensionality on the combined Direct+SOA readout}
\label{sec:Dimensionality}

To assess the role of feature dimensionality in the performance of the combined Direct+SOA readout, we construct a reduced-dimensionality representation containing 36 instead of 72 features. 
Starting from the eight available readout channels, the number of channels is reduced from eight to four by retaining two Direct and two SOA channels, with \(M=9\) temporal samples per channel.
The temporal readout window is kept fixed to that selected for the full Direct+SOA readout at each input coupling condition.
All 36 possible combinations of two Direct and two SOA channels are evaluated, and, for each input coupling condition and benchmark task, the combination yielding the lowest validation error is selected.
The resulting 36-feature Direct+SOA representation is then compared with the full 72-feature Direct+SOA readout and with the 36-feature Direct and SOA readouts.
The reduced Direct+SOA representation thus has the same feature dimensionality as either individual branch, allowing the effect of combining the two physical representations to be examined independently of feature dimensionality.

\begin{figure}[!t]
\centering
\includegraphics[width=\linewidth]{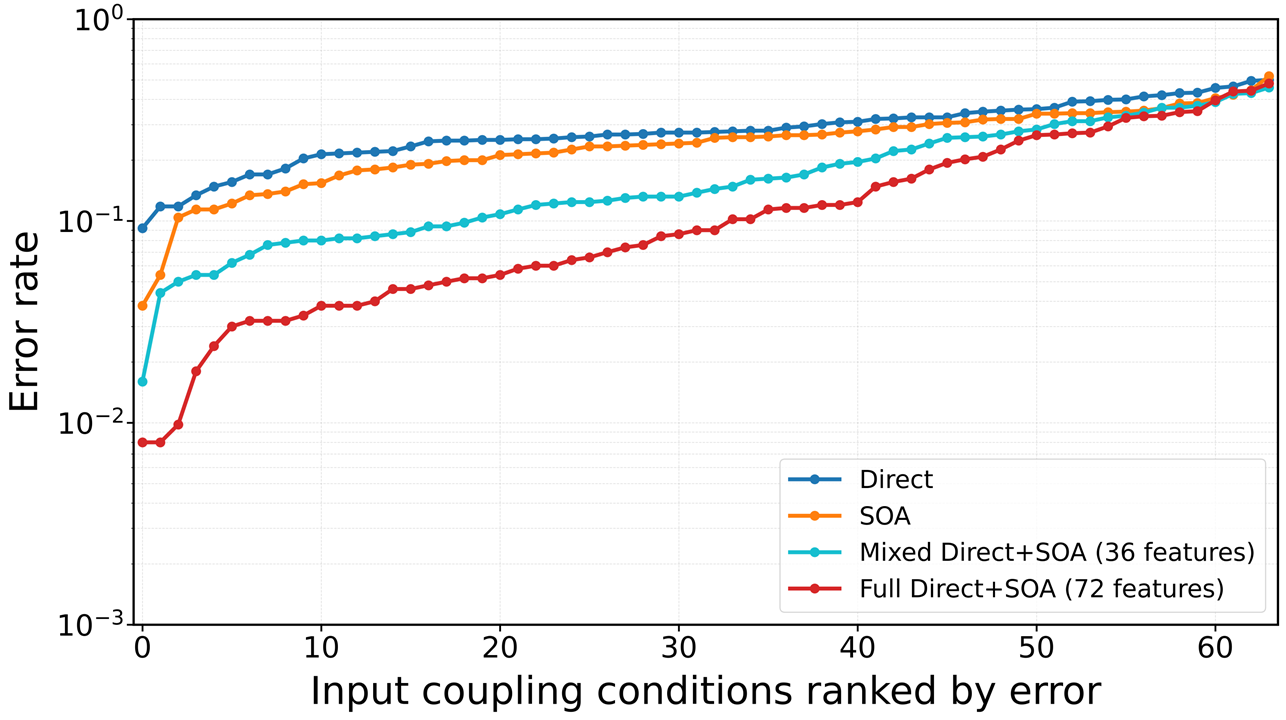}
\caption{Classification error rate for the parity(7--6--5) task using the 36-feature Direct and SOA readouts, the reduced 36-feature Direct+SOA
readout, and the full 72-feature Direct+SOA readout. For each representation, the 64 possible input coupling conditions are independently ranked by increasing classification error. 
The reduced Direct+SOA representation contains the best-performing combination of two Direct and two SOA channels for each input coupling condition. 
The full temporal readout resolution (\(M=9\)) is used in all cases.}
\label{fig:mixed}
\end{figure}

Figure~\ref{fig:mixed} shows this comparison for the parity(7--6--5) task. 
For each readout representation, the 64 input coupling conditions are independently ranked from lowest to highest classification error. 
The reduced 36-feature Direct+SOA representation achieves lower errors than either the 36-feature Direct or SOA readout over most of the ranked input coupling conditions. 
These results suggest that combining Direct and SOA-transformed representations can provide benefits beyond increasing feature dimensionality.
The full 72-feature Direct+SOA readout provides a further reduction in error, showing that retaining all eight channels provides an additional performance benefit.
The same qualitative behavior is also observed for the XOR(7--4) and XOR(7--3) tasks. 
Thus, the improvement achieved by the combined-branch readout arises from exploiting both the direct and SOA-transformed representations, as well as from the additional features available when all eight channels are retained.

\subsection{Effect of temporal readout resolution}
\label{sec:resolution}

\begin{figure}[!b]
\centering
\includegraphics[width=\linewidth]{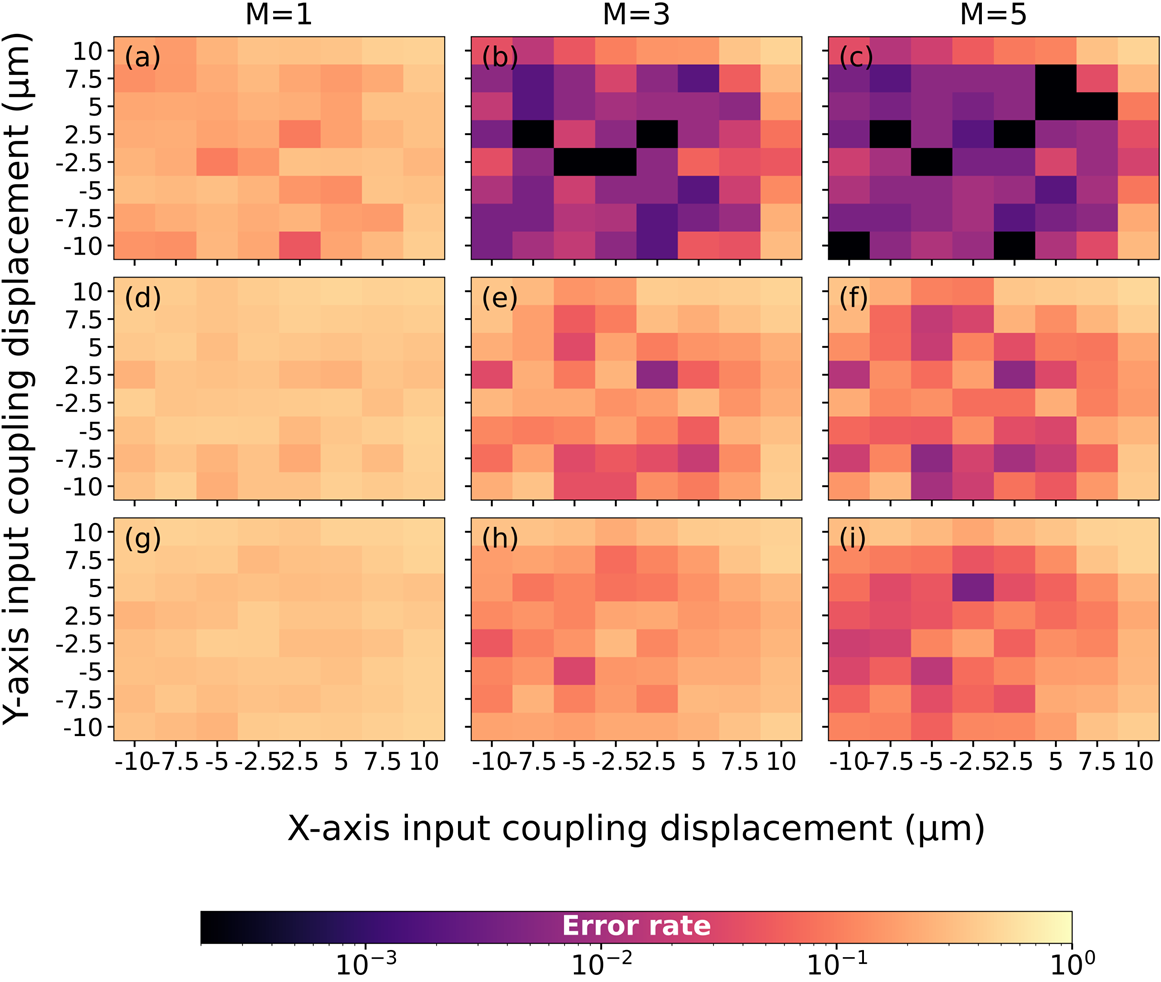}
\caption{Classification error rate across the 64 input coupling conditions for the three benchmark tasks: XOR(7--4) ((a)--(c)), XOR(7--3) ((d)--(f)), and parity(7--6--5) ((g)--(i)), using the combined Direct+SOA readout, for reduced temporal readout resolution: \(M=1\)
((a), (d), (g)), \(M=3\) ((b), (e), (h)), and \(M=5\)
((c), (f), (i)), which correspond to 8, 24, and 40 readout features, respectively. 
The indices of the temporal samples used for each value of \(M\) are shown in Table~\ref{tab:temporal_resolution}.
}
\label{fig:6}
\end{figure}

We next investigate how reducing the temporal readout resolution affects classification performance for the combined Direct+SOA readout. 
We consider \(M=1,2,3,4,5,\) and \(9\), with \(M=9\) corresponding to oversampling and the full temporal resolution used in the previous analysis. 
The sample indices retained for each case of \(M<9\), together with the resulting number of features and classification performance, are summarized in Table~\ref{tab:temporal_resolution}. 
Figure~\ref{fig:6} shows the classification error across the 64 input coupling conditions for \(M=1\), \(3\), and \(5\). 
The full-resolution case (\(M=9\)) is the one shown in Figure~\ref{fig:branch_comparison} for the Direct+SOA readout.

\begin{table}[!t]
\centering
\caption{
Summary of classification performance vs. the temporal readout resolution \(M\). 
"Samples" indicates the sample indices that are considered from the 9-sample readout of each bit. 
\(N_f\) is the resulting total number of readout features.
Each task entry reports the minimum classification error and the number of input coupling conditions (in parentheses) with an error rate of \(10^{-2}\) or less.}
\label{tab:temporal_resolution}

\setlength{\tabcolsep}{3pt}
\begin{tabular}{c c c c c c}
\hline
\(M\) & Samples & \(N_f\) & XOR(7--4) & XOR(7--3) & Parity(7--6--5) \\
\hline
1 & 5         & 8  & $7.6\times10^{-2}$ (0) & $2.3\times10^{-1}$ (0) & $2.8\times10^{-1}$ (0) \\
2 & 1,9       & 16 & $2.0\times10^{-3}$ (4) & $1.1\times10^{-2}$ (0) & $1.3\times10^{-1}$ (0) \\
3 & 1,5,9     & 24 & $\mathbf{2\times10^{-4}}$ (31) & $6.4\times10^{-3}$ (1) & $3.6\times10^{-2}$ (0) \\
4 & 1,4,6,9   & 32 & $\mathbf{2\times10^{-4}}$ (31) & $7.6\times10^{-3}$ (1) & $1.5\times10^{-2}$ (0) \\
5 & 1,3,5,7,9 & 40 & $8\times10^{-4}$ (38) & $5.2\times10^{-3}$ (2) & $\mathbf{4.0\times10^{-3}}$ (1) \\
9 & 1--9      & 72 & $\mathbf{2\times10^{-4}}$ (\textbf{40}) & $\mathbf{4.0\times10^{-3}}$ (\textbf{7}) & $8.0\times10^{-3}$ (\textbf{3}) \\
\hline
\end{tabular}
\end{table}

The results show that using only one temporal sample per bit duration (\(M=1\)) is insufficient to solve any of the three tasks, as no input coupling condition leads to low error rates. 
The performance improves substantially as additional temporal samples per bit duration are used in the classifier.
However, the temporal resolution required to achieve low-error operation depends on the task.
For the XOR(7--4) task, the error maps in Fig.~\ref{fig:6} show that much of the performance and robustness to input coupling conditions is already recovered with $M=3$, using only 24 features compared with 72 for the full-resolution case (\(M=9\)). 
For the XOR(7--3) task, low-error operation is achieved for only one input coupling condition when \(M=3\).
However, a pronounced improvement is observed as the number of temporal samples is gradually increased.
Up to seven input coupling conditions reach an error rate of \(10^{-2}\) or lower for \(M=9\). 
The dependence on temporal resolution is even more evident for the parity(7--6--5) task, for which no input coupling condition reaches the \(10^{-2}\) error level for $M\leq4$.
One has to oversample with \(M=5\) to reach an error rate of \(4\times10^{-3}\). 
Yet, the lowest error rate for \(M=5\) is not accompanied by stronger robustness across input coupling conditions.
The latter is only improved when oversampling with \(M=9\).

Overall, these results show that the maximum temporal readout resolution is not always required to achieve low classification errors. 
The 256 GSa/s sampling rate substantially exceeds the 40 GHz bandwidth imposed by the photodetection stage, resulting in partially redundant adjacent samples.
The readout dimensionality can therefore be reduced with limited degradation in performance or operational robustness.
Nevertheless, oversampling can provide useful additional features, leading in some cases to improved classification performance or greater robustness across input coupling conditions.

\subsection{Temporal position of the selected readout window}
\label{sec:localization}

\begin{figure}[!t]
\centering
\includegraphics[width=\linewidth]{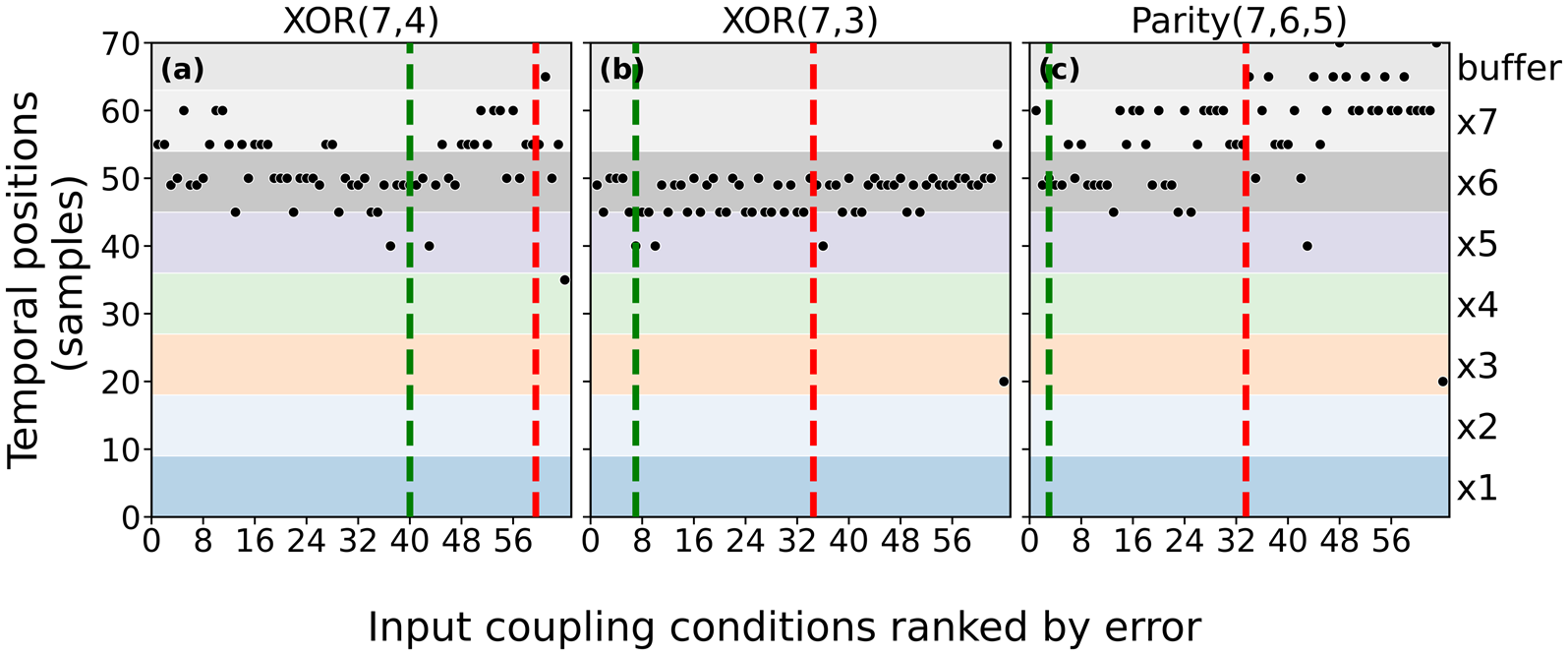}
\caption{Selected temporal readout-window positions for the combined Direct+SOA readout. 
The 64 input coupling conditions are ordered by increasing validation error along the horizontal axis. 
The black dots indicate the selected starting position of the 9-sample readout window for each input coupling condition. 
The shaded regions indicate the temporal positions of input bits \(x_1\)--\(x_7\) and the beginning of the subsequent buffer interval.
The green and red dashed lines indicate the input coupling conditions with an error rate of \(10^{-2}\) or lower and \(10^{-1}\) or lower, respectively. 
Panels correspond to (a) XOR(7--4), (b) XOR(7--3), and (c) parity(7--6--5) tasks.
} 

\label{localizations}
\end{figure}

The temporal position of the readout window can strongly affect the classification performance. 
Modal dispersion distributes the information carried by each input bit encoding across time, making the information available to the classifier dependent on the temporal region sampled by the readout.
Therefore, as described in Section~\ref{sec:task}, the position of the 9-sample readout window is optimized for each input coupling condition using the validation set of the combined Direct+SOA readout and then fixed for the performance evaluation.
Figure~\ref{localizations} shows these optimal temporal positions for the 64 evaluated input coupling conditions and the three benchmark tasks, ordered by increasing validation error.
The location of the black dots in Figure~\ref{localizations} shows that the optimal readout position depends on the task and the input condition. 
We focus on input coupling conditions with low validation error rates of \(10^{-1}\) or lower.
In particular, the XOR(7--3) task favors a relatively narrow range of temporal positions, between the samples at the end of the 5th bit and the middle of the 6th bit. 
In contrast, the XOR(7--4) and the parity(7--6--5) tasks favor later positions, up to the samples of the 7th bit. 
These differences demonstrate the importance of selecting the temporal readout position according to the task rather than using a common fixed value or range.

This task dependence of the selected position is consistent with the temporal memory generated by modal dispersion in the FMF. 
Since the different tasks require information from different combinations of temporally separated input bits, the temporal region in which these dispersed contributions become jointly accessible to the readout can also differ. 
Because of this temporal spreading, the selected window position should not be interpreted as corresponding exclusively to the bit transmitted at that particular time, since the measured response can contain contributions from multiple input bits.
Despite the variability among input coupling conditions, no pronounced systematic shift in the selected window position is observed as the validation error increases.
Overall, the selected temporal region appears to depend more on the task being performed than on the particular input coupling condition.

\section{Conclusions}

In this work, we experimentally demonstrate a dual-branch photonic architecture that extends FMF-based spatiotemporal processing to nonlinear temporal computing. 
Modal dispersion in the FMF provides high-speed temporal memory and spatiotemporal mixing, while the nonlinear SOA introduces an additional nonlinear optical transformation of the dispersed signal. 
Preserving and jointly processing the directly detected and SOA-transformed representations consistently improves classification performance and operational robustness, compared with using either representation independently. 
The results also show that temporal oversampling provides partially redundant information, allowing the readout dimensionality to be reduced.
Still, in some cases, oversampling can offer slightly improved performance. 
Overall, the proposed architecture demonstrates how different physical transformations of the same spatiotemporal signal can be jointly exploited for high-speed nonlinear temporal processing.

\section*{Acknowledgments}
The authors would like to thank Dr. Moritz Pflüger for his support with the experimental setup and configuration, and Dr. Claudio R. Mirasso, Dr. Ingo Fischer and Dr. Miguel C. Soriano for valuable discussions. This work has been funded by the European Commission under the Horizon Europe grant 101169118 (POST-DIGITAL+). It was also partially funded by the Program for Centres and Units of Excellence in R\&D, María de Maeztu, under Project CEX2021-001164-M/10.13039/501100011033.

\bstctlcite{BSTcontrol}
\bibliographystyle{IEEEtran}
\bibliography{Main}

\vfill

\end{document}